\documentstyle[12pt]{article}

\begin{document}
\def\expt#1{\mathord{<}#1\mathord{>}} 
\def\beq{\begin{equation}}
\def\eeq{\end{equation}}
\def\beqa{\begin{eqnarray}}
\def\eeqa{\end{eqnarray}}
\def\noin{\noindent}
\def\J{{\cal{J}}}
\def\V{{\bf V}}
\def\A{{\bf A}}
\def\cD{{\cal D}}
\def\bv{{\bf v}}
\def\bB{{\bf B}}
\def\bJ{{\bf J}}
\def\bE{{\bf E}}
\def\D{\Delta}
\def\pa{\partial}
\def\eps{\epsilon_{\alpha\beta}}
\def\g{{\bf G}}
\def\b{{\bf B}}
\def\Q{{\bf Q}}

\titlepage
\begin{flushright}
 QMW-PH-98-20 \\
 {\it Revised Version }\\
\end{flushright}
\begin{center} 
{\bf Orbifold Duality Symmetries  and} \\
{\bf Quantum Hall systems}\\
\rm
\vspace{4ex}
Spyros Skoulakis \footnote{email: s.skoulakis@qmw.ac.uk} and
Steven Thomas\footnote{email: s.thomas@qmw.ac.uk}\\
\vspace{7ex}
{\it Department of Physics, Queen Mary and Westfield College,}\\
{\it Mile End Road, London E1 U.K.}\\
\vspace{5ex}
ABSTRACT
\end{center}
\vspace{3ex}
We consider the possible role that chiral orbifold conformal field theories
may play in describing the edge state theories of quantum Hall
systems. This is a generalization of work that already exists in the
literature, where it has been shown that 1+1 chiral bosons  living on a
$n$-dimensional torus, and which couple to a $U_1 $ gauge field, give rise
to anomalous electric currents, the anomaly being related to the Hall
conductivity.   
The well known $O(n,n;Z) $ duality group associated with such toroidal
conformal field theories transforms the edge states and 
Hall conductivities in a way which makes interesting connections between
different theories, e.g. between systems exhibiting the integer and
fractional quantum Hall effect. In this paper we try to explore the
extension of these constructions to the case where such bosons live on a
$n$-dimensional orbifold. We give a general formalism for discussing the
relevant quantities like the Hall conductance and their transformation
under the duality groups present in orbifold compactifications. We illustrate
these ideas by presenting a detailed analysis of a  toy model based on the
two-dimensional $Z_3 $ orbifold. In this model we obtain  new classes of
filling fractions, which  generally  correspond to fermionic  edge states
carrying fractional electric charge. We also consider the relation between
orbifold edge theories and Luttinger liquids (LL's), which in the past 
have provided important insights into the physics of  quantum Hall systems.

\noindent
\newpage
\section{ Introduction}
In recent years there has been a growing body of literature which 
has revealed that duality symmetries, in their various forms, seem to play
an important role in understanding connections between the various
quantum Hall hierarchies [1]-[4]. These ideas appear to fall into two approaches. In
the first [5] (see also [6] for a more recent  discussion and 
references therein), one introduces a complex Hall conductivity 
$\tau \, = \, \sigma_{xy} + i\, \sigma_{xx} $ (where  $\sigma_{xx} \, , \, \sigma_{xy} $
are  the longitudinal and transverse  components of the conductivity tensor)
which lives in the upper half plane and naturally transforms under the 
modular group $SL(2; Z) $.  This complex conductivity enters as a
parameter of  an effective field theory that encodes localization  in the 
quantum Hall system, and which is invariant under the modular group.
On Hall plateau's, $\sigma_{xx} \, =\, 0$ 
and $\sigma_{xy} \, = \, \nu \, = p/q $ in units of
$\frac{\displaystyle e^2}{\displaystyle h} $.
 Under the assumption that the Renormalization
Group trajectories of $\tau $ ``commute'' with certain subgroups of
$SL(2;Z) $, a number of properties of both the 
 integer quantum Hall effect 
(IQHE) and fractional quantum Hall effect (FQHE) (at zero temperature)
 can be derived  [5], [6]. 
This includes derivation of critical values of  $\tau $ as well as
various universal scaling exponents.  
These results follow from the assumption that the quantum Hall phase diagram 
comes as a consequence of the action of the congruence subgroup $\Gamma_0
(2) $ on the complex conductivity $\tau $.

Another approach to the problem is based on the well established connection
between quantum Hall systems and Chern-Simons theories on the interior of
the disc [7], [8] . This equivalence holds for the particularly interesting 
Laughlin states with filling fractions $ \nu \, = \, \frac{1}{\displaystyle m} $, 
$ m \, \in \, 2 Z +1 $  and its generalizations to rational $\nu $ as well 
as to the hierarchy of Jain [3]. Because of the topological nature of
Chern-Simons theories they have observables only lying on the edge of the 
disc, which can be shown to be equivalent to a system of chiral bosons 
living on this boundary [7]-[9]. When such bosonic theories are coupled to a
background electromagnetic field, the corresponding currents have been
shown to be anomalous, with anomaly proportional to the Hall conductivity
[4], [11]-[13]  and hence leads to a natural correspondence with the boundary 
 Hall current of the bulk system. Such an identification requires the 
bosonic edge theory to be chiral, and so for $n $ such bosons, it is
natural to regard them as analogous to the coordinates of a string 
compactified on a $n$-dimensional torus $T^n$, but with additional
constraint that the underlying conformal field theory be rational.
It is now well established that toroidally compactified string theories 
possess so called $T$-duality symmetries which act on the various
compactified background fields through which the string propagates, ($G_{ij}  
$ and $ B_{ij} $ in the simplest cases), and which leave the spectrum of
states invariant. In the case of strings on $T^n $ the corresponding duality
group is $O(n,n;Z) $ [14]. 

The authors in [15] have showed how such string dualities can be used to
generate  various quantum Hall hierarchies e.g. those due to
Haldane and Jain, by choosing particular values for the
background fields which give Laughlin's fractions,  and then transforming
with certain  elements of $O(n,n;Z)$ to obtain new filling fractions $\nu '
$. Since the spectrum is invariant
under these maps, one concludes that fractions related this way have the same
spectrum which is particularly interesting. Furthermore this approach
allows a simple construction of the  wavefunctions in Haldane's hierarchy.

In this paper, pushing the stringy picture one step further, we consider
the possible role of orbifold duality symmetries in the quantum Hall effect.
Orbifolds are important kinds of compactifications in string theory, and
are obtained by taking the quotient $T^n /P $ where $P$ is some discrete
isometry  group of the torus. Effectively, taking the quotient modifies the
topology of the torus and leads to new kinds of string states [16]. At the
same time, the duality symmetries present after this process are
subgroups, in general, [18], 
of  $O(n,n;Z) $ of the torus. (Here we will assume we are dealing with
symmetric orbifolds based on underlying left-right symmetric toroidal
compactifications.) This partial breaking of toroidal duality symmetries
can be understood from the fact that there are geometrical constraints on
the background fields $ G_{ij} , B_{ij } $ etc. that ensure they allow 
the existence of the discrete isometry group $P$ . We shall see later,
it is these new
constraints that directly control  the kinds of new filling fractions one can
get following the procedures presented in [15] which we discussed above.

Before moving on, it is worth noting that  it has recently been shown  [27] that 
one can  may also find interesting new fractions by using so called mirror symmetries
of the toroidally  compactified chiral boson theory ( or its Chern-Simons 
formulation). Such mirror symmetries  are different in general,  from  the 
"T"-duality symmetries we have been discussing and their 
existence  in various string compactifications has been
well established  (see e.g. the third reference in [14] ). Indeed  mirror symmetry 
has been an important tool in understanding string compactifications on 
so called Calabi-Yau manifolds.  On the other hand,  orbifolds are well known to
encode much of the topology of such manifolds including properties such as 
mirror symmetry .  One might expect then, that the idea of using mirror symmetry
to generate new filling fractions  or make connections between those that are known, 
could also be applied to the orbifold edge models we  consider.

The paper is organized as follows. In section 2 we briefly review  the general 
formalism of orbifold edge theories obtained by quotienting $T^n $ with 
discrete isometries, concentrating on those aspects that will be relevant
to the QHE. We  show how the resulting theory may be
 coupled to a $1+1$ dimensional $U_1 $ gauge field. We shall see
 how the basic requirement
of chirality in these models can be made consistent with the quotienting
procedure, and consider the general criterion, considered in [15], for the
existence of fermionic edge states carrying charge $-e $. Implicit
equations that define the orbifold duality symmetries are also presented.
Also we discuss briefly how one can realize the orbifold construction within
the Chern-Simons framework for describing chiral bosons.
In section 3 we consider possible fermionic interpretations
of orbifold edge theories. In the case of a single chiral boson, it is
known that the corresponding fermionic picture involves the so called
chiral Luttinger liquids (CLL), [2], [21] (based on the Luttinger model [34]),
  which in turn are related to the 1+1
 dimensional Thirring model (see e.g.[17]). Generalizations of this to the case of $n$
 chiral bosons lead one to consider $n$ -flavour Thirring models with $n$ complex
 fermions coupled to a $U_1 $ gauge field.
 The orbifold edge theories we consider, should be equivalent to 
quotienting out this $n$ -flavour Thirring model by some discrete symmetry.

In section 4 we present a simple toy model based on  the two-dimensional
 $Z_3 $ orbifold. Although this does not have all the physical aspects
of quantum Hall systems we
 would like, it does illustrate some of the features that one may expect
in more general examples, whilst having the advantage of being relatively
simple to calculate with. We show how one can accommodate Haldane's  
hierarchy  $\nu \, = \, \frac{1}{\displaystyle m}\quad m \, \in \, 2 Z +1 $
within this model and then use this as a starting point to obtain other
 fractions using the surviving duality symmetries.

Finally in section 5 we end with some comments and conclusions.

\section{ Duality symmetries in chiral orbifold  edge theories }
In this section we want to review the relevant orbifold constructions that
have appeared in the string literature [16]. Although this is a large and
well developed subject, we will focus our attention on the specific issues
of how duality symmetries   manifest themselves on the zero modes of the
chiral boson edge theory, as this is the central tool we will use to
investigate the possible new filling fractions we can obtain in this system.
In this respect, we  will follow, more or less,  the notation  in [18], [19].  
Orbifold constructions can be applied in the most 
general form of toroidal compactifications, i.e. the so called Narain 
compactifications  [20] in which there are different numbers of left and right
moving chiral bosons, and in which the constant toroidal
 background fields include not just the metric and antisymmetric tensor
 fields $G_{ij} $ and $B_{ij} $ but also a Wilson line $A_{iI}$ where 
$i, j  = 1..n $ label coordinates on the torus $T^n$ and $I $ runs over the
rank of the gauge group (either $ E_8 \times E_8 $  or $ SO(32) $ ). 
It is an interesting question to ask whether  these 
more general Heterotic string compactifications, including target space
Wilson lines, have a role to play in understanding quantum Hall systems and
their possible generalizations [25]. 

In this paper we will restrict ourselves to constructing orbifolds based on
symmetric (i.e. non-heterotic) toroidal compactifications.
Following the approach of [15], our starting point is a chiral edge theory 
for  $ n$ bosonic fields $ \phi^i $ , which are taken to transform
inhomogenuously  under a local $U_1 $
gauge symmetry in  1+1 dimensions as $\delta_\lambda \phi^i =  q^i \,\lambda $, with
charges $q^i $. Their action is given by  

\beqa\label{eq1.1}
S &=&  \frac{1}{8 \pi} \int d^2 x \,  G_{ij} \cD_\mu \phi^i \cD^\mu \phi^j 
+ \frac{1}{8 \pi} \int d^2 x \,  B_{ij}\, \epsilon^{\mu \nu } \pa_\mu \phi^i
\pa_\nu \phi^j \cr
&&\cr
&& - \frac{q ^i}{4 \pi} \int d^2 x \, G_{ij} \, \epsilon^{\mu \nu} \pa_{\mu} \phi^j
A_{\nu}  - \frac{1}{4k^2} \int d^2 x \, F_{\mu \nu} F^{\mu \nu } 
\eeqa
\vskip 0.6cm
\noindent
where in eq(\ref{eq1.1}), $\cD_\mu \phi^i = \pa_\mu \phi^i  - q^i A_\mu $, 
with $A_\mu $ the 1+1 dimensional $U_1 $ gauge field, and the so called
"anomaly" term has been included  (the third term).  

The  covariant $U_1 $  current $J_{cov}^\mu $ obtained from the action $S$
is given by  
  
\beq\label{eq:1.2}
J_{cov}^\mu =  \frac{q^i}{4 \pi} G_{ij} ( \cD^\mu \phi^j + \epsilon^{\mu \nu}
\cD_\nu \phi^j )
\eeq
\vskip 0.6cm
\noindent
which has the property that it is chiral, i.e. $J^{+}_{cov}
\, \equiv \, J^0_{cov} + J^1_{cov} \,    = 0 $, 
which follows from imposing the chirality condition  
$ ( \cD_0 - \cD_1 )\, \phi^i \, \equiv \, \cD_{-} \, \phi^i = 0 $,
 with $\mu = 0,1 $
running over coordinates $t,\, x  $.
 As was noted in [15], this latter constraint is
compatible with the $\phi^i $ equations of motion, due to the 
modified action including the anomaly term. Imposing the chirality
condition is crucial to identifying  the current $J_{cov}^\mu $ with the
Hall current, at the edge of a sample   in 2+1 dimensions. Indeed, the 
Hall conductivity  $\sigma_H $ is obtained through the anomalous
conservation law of $ J_{cov}^\mu $ 

\beq\label{eq1.3}
 \pa_\mu J_{cov}^\mu  = - \frac{q^i G_{ij} q^j }{4 \pi} E  = - \sigma_H E
\eeq
\vskip 0.6cm
\noindent
where $E = \pa_0 A_1 - \pa_1 A_0 $ is the 1+1 dimensional electric field.

In the string theory approach, $\phi^i $ label compact coordinates of 
the string as it moves on the torus $T^n $, which has constant metric 
$G_{ij} $ and antisymmetric tensor background fields $B_{ij} $.
 The  
spectrum of such theories is well known to possess duality symmetries
\footnote{It should be noted however, that the gauge fields $A_\mu $ are 
absent in the string approach so that strictly speaking the presence of such
fields breaks the duality symmetries. Such a breaking is thus due to the
response of the Hall system to a electromagnetic field, rather than by the
system itself, as has been argued in [15] }
[14],
which exchange momentum states with winding states, and at the same time 
transform the background fields $G_{ij} $ and $B_{ij} $. These symmetries
therefore map the Hall conductivity $\sigma_H $ to new values in general,
and this has been exploited in [15] in making connections between various 
Hall hierarchies, and indeed to relate the integer and fractional 
QHE. We shall review some  details of these results later, when we 
discuss the situation that applies to orbifolds.

First let us consider applying the orbifold construction to the chiral
boson action (\ref{eq1.1}). The presence of gauge fields $A_\mu $ again
makes this construction not completely equivalent to the string case, and
as we shall see later there are important differences. The mode expansion
of the fields $\phi^i (x, t) $ is 

\beqa\label{eq1.4} 
 \phi^i (x,t) &=& \frac{1}{2} ( \phi_L^i (x_+) + \phi_R^i (x_-) ) \cr
&&\cr
\phi_L^i (x_+) &=& \phi_0^i + x_+ \, G^{ij}P_{Lj}  + \sum_{m > 0}
\frac{1}{\sqrt{m}} [ a_{Lm}^i e^{-i m x_+} + h.c ] \cr
&&\cr
\phi_R^i (x_-) &=& \phi_0^i +   x_- \, G^{ij}P_{Rj} + \sum_{m > 0}
\frac{1}{\sqrt{m}} [ a_{Rm}^i e^{-i m x_-} + h.c ]
\eeqa
\vskip 0.6cm
\noindent
with $x_+ \, = \, t + x $ and  $x_- \, = \, t - x $, 
and $a^i_{L,Rm} \,( a^{\dagger i}_{L,Rm} )$ the usual
 anhilation/creation operators. The left and right moving momentum
$P_{{L,R}j} $ have the expansion

\beqa\label{eq1.5}
P_{Li} &=& \frac{1}{2} (\,  2 m_i + (G-B)_{ij} \, N^j  \, )\cr
&&\cr
P_{Ri} &=&  \frac{1}{2} (\,  2 m_i - (G+B)_{ij} \, N^j \, )
\eeqa
Here $m_i  = {\rm spec }\,( p'_i ) $ is the allowed spectrum of values of the 
canonical momentum operator $p'_i $ for the chiral boson moving on $T^n $,
and $N^i $ are corresponding winding numbers. As in [15],   $m_i $ is allowed to 
assume both integral and half odd integral values.  Let us assume that
$T^n \equiv  R^n / \Lambda $ where the $n$ -dimensional lattice 
$\Lambda $ has a set of basis vectors   $e^i_a ,\, a \,= \,1...n $, so that any
lattice vector $v^i $ can be decomposed as $v^i = \sum_a e^i_a \,l^a $ with 
integer valued coefficients $l^a $. One can also define the dual lattice
basis ${\tilde e}_i^a $ which has the property that 

\beq\label{eq1.6}
{\tilde e}_i^a e^j_a = \delta^j_i  \qquad , \qquad
 {\tilde e}_i^a e^i_b = \delta^a_b
\eeq  
 
The index $i$ is often referred to as labelling the ``space'' basis, whilst 
the index $a$ labels the ``lattice'' basis [18]. Using $e^i_a $ and its dual,
one can express various quantities in either of these bases. For example
the metric on the torus  $G_{ij}  = {\tilde e}_i^a \, G_{ab} \, {\tilde e}_j^b $
etc. , where $G_{ab} $ is the toroidal metric in the lattice basis. Similar
relations hold for other fields. 

Working in the lattice basis, the well known formulae for the zero mode 
(i.e. ignoring the oscillator contributions ) parts of the Hamiltonian $H_0
$ and generator of $x$ translations,  $S_0 $,    are 

\beqa\label{eq1.7}
 H_0 &=& \frac{1}{2} ( P^t_L G^{-1} P_L + P^t_R G^{-1} P_R ) \cr
&&\cr 
 S_0 &=& \frac{1}{2}  ( P^t_L G^{-1} P_L - P^t_R G^{-1} P_R ) 
\eeqa
which can be written in a more compact form as 

\beq\label{eq1.8}  
H_0 ={1\over2}  {\bf u}^t {\bf \Xi}  {\bf u}  \quad , \qquad  S_0 = 
{1\over2}  {\bf u}^t {\bf \eta}  {\bf u}
\eeq
where
\beq\label{eq1.10} 
\bf u = \pmatrix{\bf n \cr \bf m}, \quad {\bf \eta}  = 2 \, \pmatrix{\bf 0 & {\bf I}_n
\cr{\bf I}_n &\bf 0 },\quad  {\bf  \Xi } = \frac{1}{2} \, \pmatrix{({\bf G}-{\bf B}){\bf G}^{-1}({\bf
G}+{\bf B}) & {\bf B}{\bf G}^{-1}\cr - {\bf G}^{-1} {\bf B}&  {\bf
G}^{-1}}
\eeq
\vskip 0.6cm
\noindent
${\bf I}_n$ being  the
identity matrix in $n$ dimensions, and the components of the vectors 
${\bf n}, \,  {\bf m} $ given by $N^i $ and $M_i \, \equiv \, 2m_i $ respectively.
 Note that all the background $G_{ab} $
and  $B_{ab} $ dependence lies totally in $H_0 $, and that with this choice
of basis, the quantum numbers ${\bf u} $ are integer valued.

Duality symmetries are all those integer-valued linear
transformations of the quantum numbers leaving the spectrum invariant.
Denote these linear transformations by ${\bf\Omega }$ and define their action
on the quantum numbers  ${\bf u}$ as 
\beq\label{eq1.11}
{\bf\Omega} : {\bf u} \longrightarrow {\bf S}_{\Omega}( {\bf u} ) \, = \, { \bf\Omega}^{-1}
{\bf u} 
\eeq
${\bf\Omega}$ should satisfy the constraint that 
$\bf\Omega^{t}\, \bf\eta \, \Omega= \bf\eta$ which immediately implies that 
  ${\bf\Omega}$ is an element of $O(n, n; Z)$. The backgrounds $G_{ij} $
  and $B_{ij} $ are transformed via the change
\beq\label{eq1.12} 
{\bf\Xi}\,  \longrightarrow \,  {\bf\Omega }^t \,  {\bf \Xi } \, {\bf\Omega} 
\eeq

 The orbifold is defined as the quotient
of the torus by a group of discrete  isometries, the point group $P$ 
which is usually taken to be a cyclic group. In the next
 section we shall consider a simple two-dimensional orbifold based on the 
discrete group $Z_3 $. Let $\theta ^i_j  $ be the action of an element of 
$P$ on the torus coordinate $\phi^i $, i.e.

\beq\label{eq1.13}
 P:  \phi^i \longrightarrow \theta^i_j \,  \phi^j 
\eeq
it follows that if $P$ is a discrete automorphism of $\Lambda $ then we
have 
\beq\label{eq1.14}
\theta^i_j \, e^j_a \, = \,  e^i_b\,  Q_{ab}
\eeq
where the $ n \times n $ matrix $\Q $ is integer valued and forms a representation of $P$.
From this it follows there is a natural action of $P$ on the quantum
numbers ${\bf u} $, namely, 
\beq\label{eq1.15}
P :\,  {\bf u} \longrightarrow \, {\cal {\bf R}}\, {\bf u}\, ; \qquad 
{\cal {\bf R}}\, = \, \pmatrix{{\bf Q} & {\bf 0} \cr{\bf 0}&{\bf Q}^* }
\eeq
and $\Q^* =  ({\Q^t}{)}^{-1} $.

In considering the applicability of the bosonic theory (\ref{eq1.1})
to the orbifolds,  important additional constraints have to be placed 
on the backgrounds $G_{ab} $ and $B_{ab} $. These follow from demanding
invariance of the action $S$ under the point group $P$ which yields 
\beq\label{eq1.16} 
\Q^t \,  \g \,  \Q \, = \, \g \qquad , \qquad \Q^t\,  \b \, \Q\, =\, \b
\eeq

These constraints dictate that the background fields have to have the 
right ``shape'' in order to admit a discrete isometry.
In the present context, such constraints have important consequences
regarding the possible values that the Hall conductivity $\sigma_H $ can 
take, since this depends explicitly on  $G_{ij} $, which we shall 
discuss in more detail shortly. The observant reader will have noticed that
the conditions (\ref{eq1.16}) are strictly applicable in the string
context, when there is no gauge field $A_\mu $ present. However, 
these conditions will still be applicable, and the action $S$ invariant
under the point group $P$ if we demand that under the action of $P$, 
the gauge fields $A_\mu $ remain inert, while the charges $q^i$ transform as
\beq\label{eq1.17} 
P: \, q^i\, \longrightarrow \, \theta^i_j  \, q^j
\eeq
which then implies that the quantity $\cD_\mu \phi^i $ transforms
homogeneously under the action of the point group. It can be easily seen
that as a consequence of this, together with eqs(\ref{eq1.14}) and
 (\ref{eq1.16}), 
the conductivity $\sigma_H $ after orbifolding, is just the $P$-invariant 
Hall conductivity on the torus. 

Another important point we must address is the compatibility of the chirality
condition we must impose on $\phi^i$ namely  $\cD_- \phi^i = 0 $ and the 
action of  the  point group $P$.  In particular we should hope that the
chirality condition still holds under this group action. As was shown in [15] , this chirality condition imposes 
restrictions on the allowed spectrum of zero modes, which can be written in
the compact form
\beq\label{eq1.18}
{\bf\Xi}\,  {\bf u} \, = \,  {\bf  \eta }\, {\bf u}
\eeq
That eq(\ref{eq1.18}) correctly encodes the chirality condition,  is
clear when one recalls that 
$H_0 $ and $S_0 $ are proportional to the zero mode parts of $ \langle L_0+
{\bar{L}}_0  \rangle $ and 
$ \langle L_0 - {\bar{L}}_0  \rangle $ respectively,
where $L_0$ and ${\bar{L}}_0 $ are the  (diagonal) left and right moving
Virasoro generators.

Now it follows from  the constraints imposed earlier on $G_{ab} $ and
$B_{ab} $ and the definition of
 the  twist matrix ${\bf R} $ that 
\beq\label{eq1.19}
{\bf R}^t  \, { \bf \Xi }\,  {\bf R} \, \, =  \, \, {\bf \Xi } \qquad , \qquad  
 {\bf R}^t \,  { \bf \eta } \, {\bf R} \, = \, \, \, {\bf \eta }
\eeq
Using (\ref{eq1.19}) it is easy to verify that the chirality condition is 
indeed  preserved under the action of $P$. This is an important result
since otherwise it would be hard to 
see how the orbifolding process has a role to play in quantum Hall systems,
since we know they correspond to chiral edge theories [2],[4],[12],[13],[15].

Now we  want to consider what happens to the duality symmetries in passing
from  the toroidal to the orbifold edge theory. 
 Again it is well known from the  string approach [18] that  in the process of
 orbifolding,  the toroidal duality group  $O(n,n;Z) $ is broken in general
 to a subgroup $\Gamma $ say, where  the  elements of $\Gamma $ are those
 transformations $ {\bf \Omega }$ of $O(n,n;Z) $ satisfying
\beq\label{eq1.20}
  [ {\bf R}\,,\, {\bf \Omega}]  \,=\, [ {\bf R}^2 \, , \, {\bf \Omega}]
  \, =\,  ....  \, =\, [ {\bf R}^{N-1}\, , \, {\bf \Omega} ]  \, = \, 0
\eeq
\vskip 0.6cm
\noindent
where in eq(\ref{eq1.20}) we have considered the simplified case of a point group $P $
 isomorphic to the cyclic group $Z_N $, with generator represented by ${\bf
   R } $.
 Similar conditions,  namely that the surviving duality group on the orbifold must
 commute with all the elements of  $P$ holds for  more general point groups.

 This result will again have important consequences  in determining the
 kinds of  filling fractions  and how they are related in the orbifold edge
 model.
 We shall see how this works in a specific model in the next section.

Finally in order to complete our construction, we need to identify  (as in
the toroidal case) certain  fermionic states in our spectrum, which we want
to identify with the electrons at the edge of the Hall sample. This implies
 also that among such states, we should look  for  those  with 
charge $-e $ . In [15],  it was emphasized that these additional constraints
give the correct physical interpretation of the various filling fractions
obtained
 by  specific  choices of the metric $G_{ab} $
 and those obtained by acting with duality transformations.  
In the toroidal case, the system  consists of bosons,  so fermionic states
can only be obtained through solitonic configurations via the chiral vertex operators
 $V(x_+ ) \, = \, : \, e^{i M_{Li} \phi_L^i  } \,:$, where $M_{Li} $ are
 just the allowed values of the left moving momentum $P_{Li} $. 
Note that  $M_{Li} $ need not be integer in general. 
In  passing to the orbifold, such states will also  be present  in the
untwisted sector
 (  $ V( x_+ ) $ is invariant under the point
group action since  the $M_{Li} $ also rotate under the action of $P$ ),
although in 
general there will be additional sectors of the total Fock space of states
which will correspond to the fields $\phi^i $
satisfying twisted boundary conditions. We will discuss these sectors shortly.  
In the toroidal approach, taking these various points into account, there are
three fundamental 
constraints that have to be satisfied (here we write them in the lattice basis):
\newpage
\beqa\label{eq1.21}
(i)    \qquad  ( G+ B)_{ab}\,  N^b \,   & =  & \,  M_a  \cr
&&\cr
(ii)\quad \qquad \qquad G_{ab} \, N^b \,& = &\, M_{La} \quad ,\,  M_{La'} \,  =
\, -1 \,\quad  {\rm for}\, {\rm  some}\,  a' \,
 \in \, (  1, 2, ... n ) \cr 
&&\cr
(iii)\quad \qquad \qquad  M_a \, N^a  \, &  =  &  \,  2 p +1 ,  \qquad 
 p \, \in \, Z
\eeqa
The first condition $(i) $ in  eq(\ref{eq1.21}) is just the chirality
condition we met earlier. The second condition  implies that the charge of
 the state created by $
V(x_+ ) $, $q $,  which is given by 
\beq\label{eq1.22}
q \, = \, M_{Li} \, q^i  \, = \, M_{La} \, q^a 
\eeq
is  $-e $ if we choose  $q^{a'} = e  \, , \,  q^a = 0 \, , \,
a \neq a'$. (Note $a'$ takes only one of the values $1,...n$).
   The final condition $(iii) $,
is required in order that  $V(x_+ ) $ satisfies anti-commutation relations .

Now we have to consider if these constraints are compatible with the point
group action. 
We have already seen that  condition  $(i) $ is preserved under $P$. Since
             $(iii)$ may be written in the form  $ {\bf u}^t \, {\bf \eta} \, {\bf u} \, = \, 2
 p +1  $
 then this  is also  compatible since ${\bf \eta } $ is invariant under $P$.
  Condition  $ (ii) $,  like  $(i) $ is not invariant but
 rather  both sides of the  equation transform in the same way. 
However the purpose of this condition was as we noted, 
to enforce that at least some particles in the spectrum have charge $q =
-e $. 
 But in fact we see that the charge $q $ defined in (\ref{eq1.22}) is   invariant under $P$
since we saw previously that 
it was crucial that the charges $q^a $ transform under the point group,
otherwise the action  $S$ 
will not be invariant.  Note that we are not saying that if  $(i) - (iii) $
 are satisfied  in the toroidal case, they will automatically be satisfied
 for the orbifolding. It  {\it remains}  to  find choices  of 
backgrounds  $G_{ab} $ and $B_{ab} $  compatible with the point group $P$
which explicitly
satisfy these conditions.  But once such  a solution is found, the action
of the point group will preserve it.
 
We now return to the issue concerning the existence of twisted sector states
when one couples the bosonic theory to a world-sheet $U_1 $ gauge field as 
we have described in section 1. Without such couplings, the fields
 $\phi^i ( z, \bar{z} ) $ satisfy boundary conditions $\phi^i (z\, e^{2\pi
   \, i } , {\bar{z}}\, e^{-2\pi   \,i   })
 \, = \, ({\theta^p})^i_j \, \phi^j (z, \bar{z} ) + 2 \pi v^i \, , \quad p \, =\, 0,1,..N-1 $ where $N$ is the order of the point group ( $Z_N $ in this
 example) and $v^i \, \in \, \Lambda $. The untwisted sectors correspond to
 $p \, =  \, 0 $ and satisfy the usual toroidal boundary conditions. 
The other values of $p$ define the so called twisted sectors, and they are
characterized by oscillators whose mode numbers are fractional, compared to
the integral ones defined in the expansions eq(\ref{eq1.4}),
appropriate to the untwisted sector. 

Now when one allows $\phi^i $ to transform under a local $U_1 $
gauge transformation, the gauge invariant  action (\ref{eq1.1})  implies
that in general, $\phi^i (z, \bar{z} ) $ with non-vanishing charges $q^i $ are
 not allowed to have twisted boundary
conditions as described above, because this is incompatible with its
transformations under the local $U_1 $. This can be seen from the fact that
the Lagrange density will not be a single valued function as we encircle a
point in the complex plane, which is required if it is a local density.
This is not the case when the  $U_1 $  is absent, since then
$\theta^i_j $ factors coming from the twisted boundary condition on each
$\phi^i $ cancel. Thus it is peculiar to the coupling of $\phi^i (z,
\bar{z} ) $ with
$A_\mu $, that the effect of orbifolding the toroidal theory with $Z_N $
group is to place restrictions on the various backgrounds $\g ,  \b $ ( and 
Wilson lines $ A_{iI} $ if they are present ) and not to generate twisted
sectors. The only exception to this is the case of $Z_2 $ orbifolds, since
in principle the single $U_1 $ gauge field $A_\mu $ could satisfy either
 periodic or antiperiodic boundary conditions, and this might allow twisted
 sectors to occur for non-vanishing charges, depending on how the $Z_2 $
point group is realized on $\phi^i $. Further investigations of these
possibilities would be interesting.

 Twisted sectors are allowed for fields $\phi^i $ whose charges are
zero. But such chargeless fields decouple from the current $J_{cov}^\mu $ 
and would seem not to affect directly  quantities of interest
 like $\sigma_H $ etc. However there is a subtle an interesting way in
 which they can influence what the effective duality symmetry that acts on
 $\sigma_H $ can be. This can happen in  higher dimensional orbifolds (
$ n \, > \, 2 ) $, where it is possible that  for some point groups, an element
$\theta^p $ say, has  a  fixed plane as a opposed to a fixed point,
associated with it. If we imagine going to  a block-diagonal basis then
 one has 
$\theta^p \, = \, {\rm diag} ( {\bf \theta}^{(1)} , {\bf \theta}^{(2)}, ...
 {\bf I}_2 ,..  {\bf \theta}^{(r-1)} , {\bf \theta}^{(r)} ) $, in the
 $n\, =\,  2\, r$-dimensional case, where ${\bf I}_2 $ acts on  
say the $j$th plane, and the  $2 \times 2 $ matrices $\theta^{(m)} $ act
 on separate planes. Now if the charges of $\phi^i $ were all zero except
$\phi^{i=j}\,$ and $ \phi^{i=j+1} $, then effectively we have  $2\, r - 2 $ fields 
satisfying twisted boundary conditions, whilst  the latter two fields 
just satisfy the usual toroidal boundary conditions for compactification on 
a $2$-torus.  Thus essentially from the point of view of  the fields that
 couple to
$A_\mu $ it looks like we  are back to the $n = 2 $ example discussed
earlier, where we might anticipate a $PSL(2, Z) \times PSL(2,Z)  $ duality
 group.
However this is not quite correct in general. This is because 
 the $T^2 $ being 
picked out in the above example is a subspace of the total $T^n $. How this 
embedding is realized (e.g. if  the  $n$ -dimensional lattice  $ \Lambda_n  $ 
is decomposable as $\Lambda_{n-2} \oplus \Lambda_2 $, with $\Lambda_2 $ the 
lattice associated with $T^2 $, or if  $\Lambda_n $ is not decomposable 
in this way)  will determine the actual duality symmetry acting in these
 directions. If the fixed plane $T^2 $ is associated with sublattice
 $\Lambda_2 $, then the duality symmetries are associated with $PSL(2,Z) $
otherwise it is generically a subgroup.
This property is well known from the string literature, in particular
 there
are many examples [23], [24]  where the effective duality symmetry  associated  with 
the $T^2 $  consist of so called congruence subgroups, 
$\Gamma_0(p)  $ and $\Gamma^0(p) $  of $PSL(2, Z) $  where

\beqa\label{eq1.90}
\Gamma^0 (p)  \, &  =  &\, \pmatrix{a&f \cr c& d }  \qquad  a\, d - f\, c \, = \, 1 \quad , \quad  f \, = \, 0 \,
 {\rm mod} \, p   \cr
&&\cr
&&\cr
\Gamma_0 (p) \,  & = &\, \pmatrix{a&f \cr c& d }  \qquad  a\, d - f \, c \, = \, 1 \quad , \quad  c \, = \, 0 \,
 {\rm mod} \, p   
 \eeqa
\vskip 0.6cm
\noindent
As an illustration of these ideas, table 1  lists the known 6-dimensional
non-decomposable $Z_N $
orbifolds  and their corresponding lattices.
\footnote{6-dimensional orbifolds are important in compactifications of
  10-dimensional superstring theory to $d=4$ , which is why they have
 particular relevance.  However, in the present context we can be
  more general, and  can  consider $n$-dimensional non-decomposable
  lattices}
The second column in this table gives the eigenvalues  $( \xi_1 , \xi_2 , \xi_3 ) $
 of the diagonalized point group generators
written in the form $\theta \, = \, ( {\displaystyle e}^{2 \pi i \xi_1 } , 
{\displaystyle e}^{2 \pi i \xi_2 } , {\displaystyle e }^{2 \pi i \xi_3 }) $.
In table 2  the corresponding
duality groups acting on the various $T$ and $U$ moduli associated with 
particular fixed planes are listed (see [19] for further details).   
\begin{table}
\caption{ Examples of 6-dimensional non-decomposable orbifolds and their
  corresponding lattices}
\begin{tabular}{|c|c|c|}
\hline
Orbifold &Point group generator &Lattice \\
\hline
$Z_4-a$ & $(1,1,-2)/4$& $ SU(4)\times SU(4)$ \\
$Z_4-b$ & $(1,1,-2)/4$&$SU(4)\times SO(5)\times SU(2) $\\
$Z_6-II-a$ & $(2,1,-3)/6 $& $SU(6)\times SU(2)$ \\
$Z_6-II-b$ & $(2,1,-3)/6$&$SU(3)\times SO(8)$ \\
$Z_6-II-c$ & $(2,1,-3)/6$&$SU(3)\times SO(7)\times SU(2)$ \\
$Z_8-II-a$ & $(1,3,-4)/8$&$SU(2)\times SO(10)$ \\
$Z_{12}-I-a$& $(1,-5,4)/12$&$E_6$\\
\hline
\end{tabular}
\end{table}
\begin{table}
\caption{Duality groups associated with the moduli of certain fixed planes
  in non-decomposable $Z_N$ orbifolds}
\begin{tabular}{|c|c|}
\hline
Orbifold & Duality group \\
\hline
$Z_4-a$ & $\Gamma_{T_3/2}$=$\Gamma^0(2),$\quad
$\Gamma_{U_3}$=$PSL(2,Z)$ \\
$Z_4-b$ & $\Gamma_{T_3}$=$\Gamma^0(2),$\quad
$\Gamma_{U_3}$=$\Gamma_0(2)$ \\
$Z_6-II -a$  & $\Gamma_{T_3}$=$\Gamma^0(3)$,\quad
$\Gamma_{U_3}=\Gamma_0(3)$, \quad$\Gamma_{T_1/2}$=$PSL(2, Z)$ \\
$Z_6-II -b$ & $\Gamma_{T_3}$=$\Gamma^0(3)$,\quad
$\Gamma_{(U_3+2)}=\Gamma^0(3)$, \quad $\Gamma_{T_1}$=$PSL(2, Z)$ \\
$Z_6-II-c$ & $\Gamma_{T_3}$=$\Gamma^0(3)$,\quad $\Gamma_{U_3}=\Gamma_0(3)$, \quad
$\Gamma_{T_1}$=$PSL(2, Z)$ \\
$Z_8-II -a$&$\Gamma_{T_3}$=$\Gamma^0(2)$,\quad
$\Gamma_{U_3}=\Gamma_0(2)$ \\
$Z_{12}-I-a$&$\Gamma_{T_3/2}$=$PSL(2, Z)$ \\
\hline
\end{tabular}
\end{table}

A  consequence of all this is that in edge theories corresponding to non-decomposable
orbifolds,  generally there will be a restriction on the allowed
 $PSL(2, Z) $ elements that  one may 
use to generate new Hall fractions  by acting on $\sigma_H $  
compared  to t he toroidal case  studied
in [15]. One might expect this phenomenon to generalize to the case where
some point group elements have fixed subtori $T^{2p} \, \, p > 2 $ associated with them,
in which case the duality group would be a subgroup of 
$O(p, p; Z) $. In passing, it is interesting to note that congruence subgroups
have been suggested as playing an important role in the  understanding
of various Hall hierarchies [5], [6]  so it would be worthwhile to  investigate
further orbifold models of the type discussed above. 

Before moving onto the next section, it is worthwhile recasting the
previous orbifold construction, within the Chern-Simons framework, since
this has proved a particularly important probe for studying the physics
associated with quantum Hall systems  in the past. First we remind the
reader that the compactified chiral boson edge theory, which is the
starting point for our construction, can be understood in terms of a $U_1^
n $ Chern-Simons gauge theory, with $n \, \, (2+1) $-dimensional  gauge field
 (1-forms ) $\V^i  \, \,  i = 1 ...n $ where the latter are taken to describe
 pure gauge degrees of freedom $\phi^i $. \footnote{ Here we follow the
presentation in  [28] } The $2+1$-dimensional action is 
\beq\label{eq1.23} 
K \cdot S_{CS}^{[U_1^n]} [ \V ] \, = \, \sum_{ij} \frac{K_{ij}}{8 \pi } \,
  \int_{\cal{M} }\V^i \, \wedge d \, \V^j 
\eeq
\vskip 0.6cm
\noindent
with $K_{ij} $ a real valued matrix. In [26], [27]  it was shown that the above
theory induces a conformal field theory on the boundary $\Sigma \, = \, \partial 
\cal{M} $ given by 

\beq\label{eq1.24}
 S[ \phi] \, = \, \frac{1}{8 \pi} \int_{\Sigma} d^2 x \, K_{ij} 
\,  \partial_\mu \phi^i \partial_\nu \phi^j 
\eeq 
\vskip 0.6cm
\noindent
where in eq(\ref{eq1.24}) we have used the notation introduced at the
beginning of this section. Now writing  $K_{ij} = G_{ij} + B_{ij} $ we see
that the boundary action (\ref{eq1.24}) is equivelant to the ungauged
version of the original action we started with eq(\ref{eq1.21}). In
deriving this we note that although the antisymmetric tensor term in
 eq(\ref{eq1.24}) is proportional to the $1+1 $ dimensional metric $g^{\mu
   \nu } $, in fact the difference between this term and the one involving
$\epsilon^{\mu \nu } $ in (\ref{eq1.1}) is proportional to the constraint
$\partial_- \phi^i $. Thus we see that the symmetric part of $K_{ij} $ can
be associated with the metric on a torus $T^n $ which explains 
the connection between the above Chern-Simons theory and
toroidally compactified string theory. Of course in the quantum Hall context
we are interested in the gauged version of the action (\ref{eq1.24}) as we
saw earlier. Coupling the Chern-Simons theory to
the $2+1 $ dimensional electromagnetic  gauge field 1-form $\A$ 
can be achieved in the following way [28]

\beq\label{eq1.25}  
S[ \V, \A] \, = \, K \cdot S_{CS}^{[U_1^n]} [ \V ] + \sum_i \, \int_{\cal{M}}
 \, ( \V^i \, \wedge  {\tilde{J}}_i  + \frac{q_i}{2 \pi} \, \A \, \wedge  d \V^i )
\eeq
\vskip 0.6cm
\noindent
where ${\tilde{J}}_i $ are the dual of the quasi-particle  current densities $J_i $ , 
and the second term represents a minimal coupling of $\A $ to the topological current
${\tilde{d \V}}^i$ . As was argued in [28], integrating out the $\V^i $ fields 
one finds that the corresponding filling fraction for the quasi-particles is  equal to the one 
implied by eq(\ref{eq1.3}) in the gauged chiral boson system.

Now to obtain the orbifold edge theory in this formalism,  we "mod out" the  above model with 
a discrete isometry of the torus $T^n $ whose metric is the real part of $K_{ij} $. 
The  action of this point group  $P$ on the gauge field 1-forms  $\V^i $
will be 
\beq\label{eq1.26}
 P:  \V^i \longrightarrow \theta^i_j \,  \V^j 
\eeq
Since $\phi^i $ is identified as the pure gauge degrees of freedom in $\V^i $,
this transformation of $\V^i $   implies the correct 
point group action  of $\phi^i $ we used earlier in eq(\ref{eq1.13}).

Demanding that  the ungauged action  eq(\ref{eq1.23}) be invariant with respect to this symmetry
implies that the matrix $K_{ij} $ must be $P$-invariant
\beq\label{eq:1.27}
{ ( {\theta^t})}_i^l  \, K_{lm} \, \theta^m_j \, = \, K_{ij}
\eeq
which  just reproduces the $P$-invariant conditions on the background fields $\g $ and $\b $ 
we saw earlier.  For this symmetry to extend to the gauged action  eq(\ref{eq1.25}), we require
that the charges $q^i $ and the 
quasi-particle current densities  $J_i $ transform homogeneously under the group $P$ as  in eq(\ref{eq1.17}).

Now  regarding  the possibility of  allowing for  the fields $\V^i $ to satisfy twisted 
boundary conditions  on $\Sigma $ i.e.
\beq\label{eq1.28}
 \V^i ( z \, e^{2 \, \pi \, i} , {\bar z}  \, e^{-2 \, \pi \, i}) \, =\, {( \theta^p )}^i_j \,  \V^j  (z, {\bar z})
\eeq
(where the $\V^i $  are assumed to be restricted to the boundary $\Sigma $), the situation
is really equivelant to the arguments we gave earlier when working within
the chiral boson
 formulation.
Here we see that the Chern-Simons action will not be well defined in general if we  allow
the above boundary conditions for non-trivial  matrix $\theta $. Furthermore it is the  terms
in eq(\ref{eq1.25})  that  minimally couple  $\A$ to the gauge field $\V^i$
(i.e.those 
 fields  which contribute to 
the topological currents  $ \tilde{ d \V^i} $ )  that force this conclusion upon us,  just 
as in our previous discussions this difficulty arose for  fields $\phi^i$ with non-vanishing 
charges $q^i $. As then, there are some caveats to these conclusions which could allow for 
possible  twisted sectors  for $Z_2 $ valued twists,  if the  $U_1  $ gauge field  satisfies  appropriate boundary conditions.   

Having developed a general framework which extends the previous
constructions  to the case when the chiral edge theory  involves scalars
 $\phi^i $ living
on an $n$-dimensional orbifold rather than torus $T^n  $,  later we will
consider in detail an illustrative example of the ideas
 presented here. We will see that this example does not fulfill all the physical
requirements we would like, but it has the virtues of being a simple toy model which
nevertheless illustrates the procedures involved. Given the huge 
and rich variety of orbifolds that exist, many of which have been
investigated in the string literature, it is hard to imagine that examples
cannot be found which have closer links with real quantum hall systems.
 Research  towards this end is currently underway [25].

In the next section we shall consider how  the chiral orbifold edge 
theories we have been discussing  might be connected  to  generalizations of  so called chiral Luttinger liquids, which  appear in the fermionic description of quantum Hall systems.


\section{ Connections between orbifold edge theories and generalized 
Luttinger Liquids }

A system of  many fermions at low temperatures, in the normal state, in 3d, is described
by the Landau-Fermi theory. According to the theory, there is a one to one
correspondence between the quasi-particle excitations of the interacting system and 
the excitations of the free fermion gas. The parameters of the theory are 
phenomenological and are extracted from experiments. In 1d, this Fermi-liquid 
picture breaks down [33], [35-36]. For example, the correlation functions have algebraic 
behaviour with anomalous exponents [10]. One model of interacting fermions  with universal
features is the Luttinger model [34]. In fact it  has been  proved [21]
 that the low energy
effective theory of edge excitations in the fractional quantum hall case are
described by a chiral Luttinger liquid and the anomalous exponents mentioned
above are known. 
We shall give a very brief review of some properties of LL's, based on the
recent comprehensive study presented in [17], (and references therein). 

Linearizing the free electron dispersion relation about the two
Fermi-points,
and then adding local interactions one finds a  LL 
  Hamiltonian  of the form $ H_{0} +H_{int}$, where:
 \beqa\label{eq3.1}
 H_{0} \, & = & \, \frac{\pi u_{F} }{L} \sum_{n} (\J_{n} \J_{-n} + {\bar{ \J}}_{n} {\bar{\J}}_{-n} ) \cr
 && \cr
H_{int} \, &  =  & \, \frac{\pi}{L} \sum_{n} (2 \, g_{1} \J_{n}
\bar{\J}_{n} + g_{2}\, (
 \J_{n}\J_{-n} +  \bar{\J}_n \bar{\J}_{-n}   ))
 \eeqa
where in the above Hamiltonian,  $L$ is the perimeter of the circle  which the fermions of the LL
are confined to, and $g_1 , g_2 $ are  coupling constants. $u_F $ is the Fermi velocity which enters the 
linearized dispersion relation for  the 1-d Fermi gas, $\epsilon_{lin} (k)  = u_F ( \alpha k - k_F ) $, 
with  $\alpha =  \pm 1 $ for  the right and left Fermi points  labeled  $R, L $. 
Here $k$ is the wave number and  $k_F $ the  Fermi wave number.

The  operators  $ \J_n $ and  $\bar{\J}_n $ are  the moments   of  two commuting $U_1 $ currents 
 $\J (\sigma ) $, $ \bar{\J} ( \sigma ) $

\beqa\label{eq3.2}
 \J_n \, & =  &\, \int^L_0 \, \J( \sigma ) \, e^{+2 \pi i n \frac{\sigma}{  L} } \, d \sigma \cr
 && \cr 
\bar{\J}_n \, & =  &\, \int^L_0 \, \bar{\J}( \sigma ) \, e^{- 2 \pi i n \frac{\sigma}{ L} } \, d \sigma 
\eeqa
\vskip 0.6cm
\noindent
These currents are in turn, related to  the  L, R components of a Dirac fermion fields 
$\Psi_R (\sigma ) $ and  $\Psi_L (\sigma ) $  that describes the 
infinite Dirac sea that  is present in the  infrared limit of the LL, 

\beq\label{eq3.3}
 \J (\sigma ) \, = \, : \Psi^{\dagger}_R ( \sigma )  \Psi_R ( \sigma ) :  \qquad , \qquad 
\bar{\J} (\sigma ) \, = \, : \Psi^{\dagger}_L ( \sigma )  \Psi_L ( \sigma ) :
\eeq
\vskip 0.6cm
\noindent
The operators $\J_n  , \bar{\J}_n $ satisfy the usual $U_1 \times U_1 $ 
affine algebra. 

The total Hamiltonian, in the form  eq(\ref{eq3.1}), is not diagonal in the 
current operators. This can be remedied by a Bogoliubov transformation
which redefines the currents [10]  and  in terms of  new current operators
$J_n  , {\bar{J}}_n $ the Hamiltonian  can be written in the diagonal form
\beq\label{eq3.4}
 H_{tot} \, = \, \frac{\pi u_S }{L} \, \sum_n
 (J_{n} J_{-n} + \bar{ J}_{n} {\bar{J}}_{-n} )
\eeq
\vskip 0.6cm
\noindent
with $u_S $ the modified Fermi velocity
 
\beq\label{eq3.4b}
u_S \, = \, \sqrt{ ( u_F + g_2 )^2 - g^2_1  }
\eeq
\vskip 0.6cm
\noindent
The current and density  fields  $j(\sigma ) $ and  $\rho (\sigma ) $ of the 1-d model are
identified as  [10] 

\beq\label{eq3.5}
\rho (\sigma )  \, = \, \alpha^{-1/2} ( J (\sigma ) + \bar{J} (\sigma ) )  \qquad , \qquad 
j ( \sigma ) \, = \, u_S \, \alpha^{-1/2} ( J (\sigma ) - \bar{J} (\sigma ) )
\eeq
\vskip 0.6cm
\noindent
where
\newpage
\beqa\label{eq3.6}
u_N \,  & =  & \, u_F + g_1 + g_2 \cr 
u_J \,  & =  & \, u_F - g_1 +g_2  \cr 
\alpha  \,  & = & \,   \sqrt{ u_N / u_J } 
\eeqa
$ u_N $ and  $u_J $ being the charge and current velocities of the LL.  Note athough a slight 
misuse of notation, we use the same
symbol $\alpha $  in  eq(\ref{eq3.6}) as appears in the linearized dispersion relation  we saw earlier in the non-interacting case, since it corresponds  to the same quantity   generalized to the interacting case.
Basically  the physical properties of  LL are controlled by the two parameters 
$u_S$ and  $ \alpha $. 

Now  an important feature of  the LL  Hamiltonian in its diagonal form  eq(\ref{eq3.4})
is that it can be given a CFT interpretation, at least in the infrared limit. The current -current 
form of $H_{tot} $ is suggestive of  a  Lagrangian description in $1+1 $ dimensions  which has 
4-Fermi interactions, since we have seen that each current is bilinear in terms of  the R,L
  components of  
Dirac fermions. The  Lagrangian description is just the  1-flavour massless Thirring model based on 
the group $U_1 $, with Lagrangian  (in Euclidean form space)

\beq\label{eq3.7}
{\cal L} \, = \, i\Psi^{*}_L   \partial_u \Psi_L +  i\Psi^{*}_R   \partial_v \Psi_R
  - h \Psi^{*}_L  \Psi_L \Psi^{*}_R  \Psi_R
\eeq  
\vskip 0.6cm
\noindent
where  $u \, = \, \tau - \sigma \, , \, v \, = \, \tau + \sigma $  and  $h$ the  4-Fermi coupling . 
\footnote{Here we follow the notation of  [29] }
Clearly the coupling $ h $ must be related to the parameter  $\alpha $ that 
controls the coupling strength in the LL.  We shall discuss the precise relationship shortly. 

An interesting property of the massless Thirring model is that it has been shown to be 
equivelant to  a single bosonic field $\phi (\tau, \sigma ) $ compactified on a circle of radius  $R$
[29].  This  relationship is somewhat subtle as  it turns out that to reproduce the full spectrum
of the compactified boson, the Thirring fields $\Psi_L , \Psi_R $ have to satisfy twisted 
boundary conditions, in general, in the compact coordinate $\sigma $. (Note that this twisting of 
the "fermion" boundary conditions  is not to be confused with the twistings related to 
orbifold compactifications which we are discussing in this  paper, which have to be considered on top of these.)  Since  the radius $R$ is  effectively the only parameter of the  boson model, it must be related to the coupling $h$. In [29] it was proven that complete equivalence of spectrum in the two formulations
requires  $h = \frac{1}{\displaystyle 2} ( R - \frac{1}{\displaystyle  R}  ) $, so  that $R  \, =  \, 1 $
corresponds to the usual free-fermion point.    

In [17] it was also shown how the LL has a bosonic formulation, again as one might anticipate,
the boson is  compactified on an circle whose radius  $R$ was shown to  be given  by 
$R =  \sqrt{\alpha}  $, so that  $h = \frac{1}{\displaystyle 2} (
\sqrt{\alpha}  -
 \frac{1}{\sqrt{\displaystyle \alpha }} ) $.
This  is consistent  since in the  case  of a non-interacting LL,  $u_N \, = \, u_J  $ and  so  $ R = 1 $

Having reviewed  some  basic properties of  standard  LL's,   what remains
is to discuss their
 relation to 
incompressible Hall fluids.  Basically the idea is that if one considers such a fluid  defined on an annular geometry, or equivalently on a cylinder,  then the edge theories  along each of the two boundaries 
can be described by the two chiral sectors of the Luttinger CFT.  That such an identification can reproduce the 
physical properties of Hall fluids can be understood  from studying charge
transport between the two boundaries in the presence of an adiabatic
magnetic field (see [17]).  In doing so one can learn that 
basic features  like  the transport of an integer or fractional  number of
electrons from one edge to the other  as one increases the magnetic field
by  multiples of the flux quantum $\Phi_0 $ , 
 can be predicted from the Luttinger CFT . In this way one finds
  that the filling fraction (which is related to $\sigma_H $ in the
  usual way) is given in
 terms of  the parameter $\alpha $  (or the radius $R$) of the LL by
 $ \nu \, = \, \alpha^{-1} \, = \, R^{-2} $. 

At this point we can make contact with the  (gauged) chiral boson
formalism for studying  quantum Hall
systems   [15] upon which the present paper is
based. The additional features that these  models incorporate,
compared to the bosons we have described above  is chirality and coupling
to a $U_1 $ gauge field.  But both these can be  included  in the fermionic
picture of LL's. Chirality of the boson theory means focusing on one of
the two chiral sectors of the LL and gives rise to  so called chiral
Luttinger liquids or CLL's . They are examples of   rational LL's, that is the parameter
$\alpha $ must be rational which immediately follows  since the corresponding radius of the compactified boson must also be rational to be consistent with chirality [15].  The second feature, namely coupling to an external electromagnetic field  can also be achieved (see [17] and refs. therein).  

It is interesting to note  that  the predicted relation between  the radius $R$ and  $\nu $   as seen above is the inverse of the prediction we get from the anomalies argument in the gauged chiral boson which 
gives $ \nu \, = \, R^2 $. However since there is a duality  symmetry  $ R \rightarrow 
\frac{1}{\displaystyle R} $ in the system, the two predictions are equivelant.

Now  we consider how the orbifolding procedure we have studied in
this paper,  manifests itself 
within the fermionic LL picture described above. Clearly  one needs to consider the generalization 
to $n$ bosons,  which naturally leads one to  $n$-flavour LL models, which would 
have Hamiltonian similar to those considered in eq(\ref{eq3.1}) or eq(\ref{eq3.4}) with inclusion of  an
index $i = 1....n $ on all the operators. There would be  a correspondingly larger number of 
possible coupling constants analogous to $g_1 $ and $g_2 $. The resulting model would presumably
be related in the infrared to  the $n$-flavour massless Thirring model which can be written as 
\beq\label{eq3.8}   
{\cal L} \, = \, i\Psi^{*i}_L   \partial_u \Psi^i_L +  i\Psi^{*i}_R   \partial_v \Psi^i_R
  - h_{ij} \,  \Psi^{*i}_L  \Psi^i_L \Psi^{*j}_R  \Psi^j_R
\eeq
\vskip 0.6cm
\noindent
with a matrix  of coupling constants $ h_{ij} $.  As shown in [29],  by assigning suitable boundary conditions to the fields $\Psi^i_L  , \Psi^i_R$ the various charge sectors of the model can 
reproduce the states of  a set of bosons  $\phi^i $   identified as coordinates  on a torus  $T^n $,
in the presence of metric $G_{ij} $ and antisymmetric tensor  $B_{ij} $, which was our starting point in 
section 1.  The explicit connection between these background fields and  the couplings $h_{ij} $
is  
\beqa\label{eq3.9}
h_{ij} \,  & =  &\, \frac{1}{4} {\cal B}_{ai} \, {\cal D}_{aj}  \cr 
&&\cr
{\cal B}^{ai} \, & = & \,  Z^{ai} - {\tilde{Z}}^{ai}  + Y^{ab} Z^{bi}\cr
&&\cr
{\cal D}^{ai} \, & =  & \,  Z^{ai} - {\tilde{Z}}^{ai}  - Y^{ab} Z^{bi}
\eeqa
where the real matrices $Z^{ai}  \, ( Y^{ab} ) $ are symmetric (antisymmetric) respectively 
and $\tilde{{Z}}^{ai} $ is the inverse of $Z^{ai} $ . These are related to the 
backgrounds $G_{ij} $ and $B_{ij} $ in the bosonic model, however, as discussed in 
[29], there is no unique dependence of  $Y_{ab} $ and  $Z^{ai} $ on these backgrounds. Rather there are various (equivelant) choices,  each of which determines the boundary conditions on the fermionic operators, and indeed whether the latter
are to be regarded as free fields or not.  As a  specific example, (see [29] for details)
one such choice is $ Z^{ai} \, = \,  e^{ai} \quad, \quad  Y_{ab} = \, B_{ab} $
in which case the fermions are interacting ($h_{ij} \neq 0 $ ).
It is clear from  this that  all the constraints that we must impose on the  couplets 
$(\g, \b ) $ in order to capture the correct physics of quantum Hall systems, as given 
in  (i) - (iii) of eq(\ref{eq1.21})  can now be viewed as constraints on the 
corresponding $n$-flavoured LL and  its CFT  representation eq(\ref{eq3.8}).
In addition we  need to impose the geometric constraints eq(\ref{eq1.16}) to 
encode the orbifold constraints. This will impose the following conditions on
the Thirring couplings $h_{ij} $
\beq\label{eq3.40}
h_{ab} \, = \,  {Q^t }^c_a \, h_{cd} \, Q^d_b
\eeq
\vskip 0.6cm
\noindent
 where we have written quantities in the lattice basis. Thus an immediate 
consequence of considering orbifold edge theories  to describe Hall plateau's
is that there are  further restrictions  on the  allowed form of the 
LL Hamiltonians compared to the toroidal case.

In the present formulation, we can address the issue concerning possible new boundary conditions  that 
the  various fields satisfy  in the orbifold case, (i.e.the twisted sectors).
Our previous arguments concerning this apply here, i.e. that because of the coupling to the 
$U_1 $ gauge fields it is difficult to see how general $Z_N , \, N > 2$ twisted boundary conditions 
can be allowed  for  the bosons $\phi^i$.  Note that the bosonization  rules

\beqa\label{eq3.41}
\partial_u \phi^a \, & =  & \,  \frac{1}{4}  B_{ai} \,  \Psi^{*i}_L \Psi^i_L \quad  =  \quad J^a_L \cr
&&\cr
\partial_v \phi^a  \,& = &\,   \frac{1}{4}  D_{ai} \,  \Psi^{*i}_R \Psi^i_R \quad = \quad J^a_R
\eeqa
\vskip 0.6cm
\noindent
allow for arbitrary twistings of the  $\Psi^i_{L,R} $ boundary conditions, since the
corresponding phase factors  just appear as global $U_1^n$ transformations
which leave the currents in eq(\ref{eq3.41}) unchanged.  The same is true even if one couples the Thirring model  to a $U_1 $ gauge field (which  is equivelant to the gauging that gives the corresponding
bosonic model), since this coupling is through the currents themselves. 
As we mentioned before, these twistings are independent of those  normally associated with 
orbifolds.  Orbifold  boundary conditions would imply that the currents  $J^i_{L,R}$
satisfy (in the $Z_N $ case)

\beq\label{eq3.12}
  J^i_{L,R} (\sigma + 2 \pi ) \, = \,  ( \theta^p )^i_j  \, J^j_{L,R} (\sigma )
\eeq
\vskip 0.6cm
\noindent
for $p \, =\, 1..N  $. 
In the ungauged theory,  such boundary conditions are permissible if the 
couplings $h_{ij} $ satisfy the constraint eq(\ref{eq3.40}) and the resulting 
theory would be equivelant to the bosonic orbifold spectrums  (again ungauged) we have discussed in 
the last section. \footnote{Details of this equivalence in the  $n \, = \, 1 $ case (where the orbifold is 
the line element  
$S^1 / Z_2 $) was shown some time ago [22]} 
Including the coupling to a $U_1 $ gauge field we are lead to the same conclusions 
discussed in the last section, here seen in the fermionic picture. 

Because of the equivalence of the $n$-flavour Thirring model and  bosons  on $T^n $, the duality 
symmetry group $O(n,n ; Z) $ of the latter is also a symmetry of the
former,
 and should also 
be a symmetry of the corresponding  generalized LL. The same will be true
of the orbifold model
for the  surviving duality subgroup of $O(n,n ;Z) $ discussed in the last section. The  LL picture offers an interesting alternative way of seeing how  duality acts on the filling fraction $\nu $ of  
the corresponding Hall systems, and it would be particularly interesting to investigate the details 
of this [25]. For example, determining the effect of such dualities on charge transport  from one edge to the  other  within LL would offer a physical picture of how dualities are linking states of different filling fractions, in both the toroidal and orbifold models.  

Finally, in ending this section, we make some brief comments about the finite temperature case.
It has been shown in the past [30] that using CFT models to
describe finite temperature Fermi liquids, in the Euclidean time
formalism,  leads naturally to considering CFT's defined on a 2-torus rather
than the complex plane. In CFT and in string theory one is used to the idea
that for consistency one has to put constraints on the model one is
considering by demanding that the 1-loop partition function, which is
related to the Euclidean partition function of the Fermi liquid, be
invariant under the action of the modular group of the torus  $\tau  \,
\rightarrow \, \frac{\displaystyle  a \tau + b} {\displaystyle  c \tau + d
  } $  where $\tau $ is the
complex modulus of the 2-torus. In string theory this is necessary for 
unitarity and to avoid global world sheet anomalies, since such modular
transformations are just large coordinate transformations.
These worldsheet anomalies  are related to various space-time anomalies.
 However in 
models where there is no space-time interpretation, imposing modular
invariance may not always be appropriate. 
In [17] it has been argued that for LL's corrresponding to the single boson case,  
 the appropriate choice of boundary
conditions on the world sheet torus, and their relative weight,
 is not dictated by modular invariance but by matching conditions with the
 LL partition function  (a general class of such partition functions, based on 
extended characters, has been 
calculated in [32]). This might also  be the case for our
 orbifold edge theories; the correct weights of untwisted/twisted sector
 boundary conditions on the torus should reproduce the ``twisted''
 generalized LL partition one gets by orbifolding the Thirring model.
That these weighted boundary conditions would  differ, in general,  from the modular
invariant ones, means for example that  (unlike in the usual orbifold constructions  [16]), 
 we  may  not be forced to project onto  $P$-invariant states.

\section{ Quantum Hall fractions in the $Z_3 $ orbifold: A toy  model }

In this section we will  study in detail the two dimensional orbifold $T^2
/ Z_3 $ where  $T^2 $ is the 
maximal torus associated with the group $SU(3) $.   A basis of lattice
vectors is given  by the set of simple roots ${\bf e}_1, {\bf e}_2 $  which transform 
under 
 the  $Z_3 $ point group generator $\theta^i_j , \quad   i ,j  \, =  \,
1, 2  $ as [31]
\beq\label{eq2.2}
{\bf \theta}  \, {\bf e}_1 \, = \, {\bf e}_2 \quad , \quad  {\bf \theta} \, {\bf e}_2 \, = \, - {\bf e}_1 -{\bf e}_2 
\eeq
which can easily be checked to give the $\Q$ matrix
\beq\label{eq2.3}
{\bf Q} \, = \,  \pmatrix{ 0&1\cr -1&-1}\qquad , \qquad {\bf Q}^3 \, = \,
{\bf I}_2
\eeq
Now  the  $Z_3 $ invariant  forms of the metric and antisymmetric tensor
fields are easily calculated to be 
\beq\label{eq2.4}
\g  \, = \, \frac{1}{2} \pmatrix{ 2 g_{11} & g_{11} \cr 
  g_{11} & 2 g_{11} } \qquad , \qquad
\b   \, = \,  \pmatrix{ 0 & b\cr -b & 0 }
\eeq
where $g_{11} $ and $ b $ are real parameters.
The surviving  duality symmetries acting on these fields can be  determined
as follows.
Firstly, for the theory defined on $T^2 $, the duality symmetry  group is
$O(2,2; Z)$ which is isomorphic (up to additional $Z_2$  factors) to the group
 $PSL'(2, Z) \times PSL(2,Z)
$. Each of these $ PSL(2,Z) $  factors acts on the 4-dimensional
vectors  ${\bf U}^t  \, = \, (  N^1,  N^2,  M_1,  M_2 ) $ as follows:
\beqa\label{eq2.5}    
PSL'(2,Z) \qquad  : \qquad   {\bf \Omega}^{-1} \, & = & \, 
 \pmatrix{ d' \, {\bf I}_2 & -c' \, {\bf L} \cr b' \,{\bf L} & a' \, {\bf
     I }_2  }
 \cr
&&\cr
&&\cr
PSL (2,Z) \qquad  : \qquad   {\bf \Omega}^{-1} \, & = & \,  \pmatrix{ {\bf  M} &
  {\bf  0}
 \cr  {\bf  0} &
{\bf  M}^*  } \cr
&&\cr
&&\cr 
{\bf M} \, = \, \pmatrix{ a & -f\cr -c& d } \qquad ,  && \qquad  {\bf  L} \,
= \,
 \pmatrix{0&1\cr -1&0}
\eeqa
\vskip 0.6cm
\noindent
where the integers $a,f,c,d$ and $a',b',c',d' $ satisfy $af -bc\,  = \,
a'd'-b'c'\, =\,  1 $.
Now since the matrix ${\bf R}$ in this example generates a subgroup of the $ PSL(2,Z)
$ defined in 
(\ref{eq2.5}), the unbroken  duality group $\Gamma $ in the $Z_3 $
orbifold, as determined by the conditions (\ref{eq1.20})   is  just
$PSL'(2,Z) $.
\footnote{ Strictly speaking the surviving duality group 
defined by (\ref{eq1.20} ) is $PSL'(2,Z) \times Z_3 $.
However, the  additional $Z_3 $ symmetry is not relevant 
for the purposes of this paper, and does not act on the $T$ 
modulus as defined above in (46). Indeed this $Z_3 $ symmetry is a
 subgroup of the $PSL(2, Z) $ that acts on the  complex
 $U$- modulus which appears in string compactifications on $T^2 $, given in
 terms of the metric by 
$U \, = \, \frac{1}{g_{11}} \, ( g_{12} + i \,  \sqrt{det G}  )  $.
When one passes to the orbifold and uses the $Z_3 $ invariant form of the
metric (44) , it is apparent that this $U$ modulus is just a fixed complex
constant which plays no role in what follows.}

 Defining the complex modulus 
$T \, =  \,   2 \, (\,  b +  i \frac{\sqrt{3}}{\displaystyle 2}  g_{11} \, )  $, this
surviving $PSL(2,Z) $ acts as 
\beq\label{eq2.6}
 T \, \rightarrow \,  \frac{a' \, T +  b'}{ c' \, T  + d' } 
\eeq
\vskip 0.6cm
\noindent
and is the familiar T-duality  present in  compactified string theory.

Now let us consider solving the constraints $(i) - (iii) $ given in
 eq(\ref{eq1.21}). We will begin with the choice $M_{L1} \, =  \, -1 $ and
consider alternatives later. One can  find the following expressions for
$N^2  , M_1 ,  M_2 $ and $M_{L2} $ as functions of  the  parameters  $g_{11} , b  $ and
$N^1$
\beqa\label{eq2.7}
N^2 &\,  = \,  & - 2 \, {g^{-1}_{11}} - 2 \, N^1 \cr
&&\cr
M_1 & \, = \, & -1 - 2 \, N^1 \, b - 2 \,{g^{-1}_{11}} \, b  \cr
&&\cr
M_2 & \, = \, &  -\frac{3}{2} \, g_{11} N^1 - 2 - N^1 \, b \cr
&&\cr
M_{L2} & \, = \, &  -2  - \frac{3}{2}\, g_{11} N^1 
\eeqa

From which we learn that $ g_{11} \, = \, 2/p \, ,\, p \in
 Z $ and $N^1 \, = \, \alpha' \, p \, , \, \, \alpha' \, \in \, Z $,
 and $b \, = \ k/p \,  , \,  k \, \in \,  Z $. 
Substituting these into the constraint $ (iii) $, one finds that 
\beq\label{eq2.8} 
N^1 \, M_1 + N^2 \, M_2 \, = \, 2 \, \alpha' \, p + 2 \, p \, ( 2 \, \alpha'
+ 1 )  + 6 \, \alpha'^2  \, p
\eeq
so that in fact, with the choice $M_{L1} \, = \, -1 $  the solution of $
(i) $ and  $ (ii) $  describes {\it bosonic } rather than fermionic states,
since the spin factor in (\ref{eq2.8}) is even and not odd as we require.
Given that this is not quite what we want, are there other solutions? 
We can instead impose that $M_{L2} \, = \, -1 $ (there is no exchange
symmetry in eq(\ref{eq1.21}) between $M_{L1} $ and  $M_{L2} $ so 
in principle, this  choice leads to different solutions.) Getting a charge
$-e $ state would follow from choosing $q^{i=1} = 0 \,  , \,  q^{i = 2}
\, = \, e $, whereas previously we would have $ q^{i=1} = e \,  , \,
q^{i=2} = 0 $.
In this case one finds
\beqa\label{eq2.9}
N^2   &  \, = \, &  - g^{-1}_{11} - \frac{1}{2} N^1\cr
&&\cr
M_1  & \, = \, & \frac{3}{4} \, g_{11} N^1 - \frac{1}{2} - b\,  g^{-1}_{11} -
\frac{b N^1}{2} \cr
&&\cr
M_2 & \, = \, & -1 - N^1 b \cr
&&\cr
M_{L1}  & \, = \, & - \frac{1}{2} + \frac{3}{4} \, g_{11} N^1 
\eeqa
from which we learn that, writing $N^1 = s\,   \in  \, Z $,
that  $g_{11} \, = \, 2/t ,\quad t \, \in \,  Z $ with the condition that
 $ s + t \, \in \,  2 \, {\cal Z} $. At the same time $t$ must be a
 divisor of $s$ so we write $s = \rho \, t $ for some  $\rho  \, \in  \,
  Z $ and then we find that the allowed values of $b $ are $ b \, = \,
 \frac{\displaystyle k'}{\displaystyle t} $. Finally calculating the spin term $N^t M $ we find after
 substituting the values $ N^1 = \rho \, t \quad, \quad N^2 = -\frac{\displaystyle t}{\displaystyle 2} (
 \rho + 1 ) \,  , \quad M_1 \, = \, \frac{1}{2} ( \rho (3-k') - 1  -k' )
\,  , \quad M_2 \, = \, -1 - k' \rho  $ ,
\beq\label{eq2.10}
N^t M \, = \, \frac{t}{2} \,  ( 3\, \rho^2 + 1 )
\eeq
It would appear that demanding $N^t M $ be an odd integer has two solutions
\beqa\label{eq2.11}
(i)  \qquad \qquad  \qquad 3 \, \rho^2 +1 &= & 2\, u
  \qquad t,\,  u \,  \in \,  2 Z + 1 \cr 
(ii) \qquad\qquad \qquad \qquad t \, & = & 2\, m  \qquad m \, \in  \quad 2\,
  Z+1  \qquad 
\rho \, \in  \quad 2 \, Z
\eeqa

However it is  impossible to find integers $ \rho $ and odd
integers $ u $ satisfying eq(\ref{eq2.11} )(i) so we will assume the second
solution (ii). 

We note that using this solution, we have the further condition that $ k'
\, \in 2\, Z +1 $. Thus by imposing the condition $M_{L2} \, =
\, -1 $ we have found the necessary conditions on the various parameters to
provide a solution to  $(i) - (iii) $ given in
 eqs(\ref{eq1.21}). 
Note  that with this choice of $\rho $,  one has $M_{L2}  \,
\in  \,  Z $.
 
It is easy to see that with the above choice of charges $q^i$ and metric
component $g_{11} $, the 
Hall conductivity $\sigma_H $ and the corresponding filling fraction, $\nu
$ are given by
 
\beq\label{eq2.12}
\sigma_{H} \, = \, \frac{e^2}{2 \pi} \, (\,  \frac{2}{t}\, )\qquad ,
   \qquad \nu =
\frac{2}{t} \, = \, \frac{1}{m}
\eeq
\vskip 0.6cm
\noindent
As discussed in [15] there is a condition that the inverse metric
$\g^{-1} \, \in \, 
Z $ in order that, for example, the edge state wavefunctions have
correct
 analyticity properties.
Since we have, in our example,
 
\beq\label{eq2.13} 
\g^{-1}  \quad = \quad   \frac{1}{3} \, 
\pmatrix{ 2t & -t \cr - t &  2t }
\eeq
\vskip 0.6cm
\noindent
we learn that $t$ must be integer multiples of  $3$. The fact that  $t/3 $
must be integer can be traced to the fact that we are considering a $Z_3 $
 orbifold in the present model, but in general we can expect that 
for $Z_N $ orbifolds, or those with  point groups containing $N^{th}
$-order elements,  a similar
constraint will hold involving $t/N $.  This is a particularly interesting
and clear difference between
the torus and orbifold  edge state theories,  since in the former there are
no such constraints on the 
integer appearing in the denominator of the filling fractions $\nu $, other
than it being  odd.

Having found at least one set of values of the fields $ b , g_{11} $ and
 corresponding   filling fractions 
that  define  chiral edge theories which contain fermionic particles with
charge $-e$,  our strategy, is to see if we can generate new filling fractions by
making use of the 
surviving duality group  $\Omega $ present after the orbifolding
procedure. In particular the  most stringent conditions on whether or not this can be
 achieved, will  comes from demanding that the 
transformed  "couplet" $( \b', \g' ) $   satisfy  conditions  $ (i) - (iii) $
of eq(\ref{eq1.21}).

Since in our simple example,  $\Gamma = PSL(2, Z) $ we can in fact
calculate explicitly the most general
form of the  new filling fractions obtained in this way.  From the
transformation properties of the $\Xi $ matrix defined earlier, we have
 
\beqa\label{eq2.14}
( {\g}^{-1} ) ' \quad   =  \quad & {\bf A}^t_2 &  (\, ( \g - \b ) {\g}^{-1}
 ( \g + \b ) \, {\bf A}_2 + \b {\g}^{-1} {\bf A}_4 \, )\cr
  &+& {\bf A}^t_4  \, ( - {\g}^{-1} \b {\bf A}_2 +
{\g}^{-1}
 \, {\bf A}_4  \, ) \cr
&&\cr
{\b}' \,  ( \g ^{-1} )'  \quad = \quad & {\bf A}^t_1 & \, ( \, (\g - \b )
{\g}^{-1}(\g + \b ) {\bf  A}_2 + {\g}^{-1} {\bf A}_4 \, )\cr
& + & {\bf  A}^t_3 \, ( - {\g}^{-1} \b {\bf A}_2 +
{\g}^{-1}{\bf A}_4 )
\eeqa
where the $2 \times 2 $ matrices  ${\bf A}_1 ....{\bf A}_4 $ are given in terms
of the general element $\omega  \, \in  \, PSL(2,Z) $ as
 
\beq\label{eq2.15}
 \omega \quad = \quad \pmatrix{{\bf  A}_1&{\bf  A}_2\cr{\bf  A}_3 &
 {\bf  A}_4 } \quad \equiv \quad 
\pmatrix{ \beta {\bf I}_2 & - \gamma {\bf  L} \cr \delta {\bf  L } & \alpha {\bf  I}_2 }
\eeq
\vskip 0.6cm
\noindent
with  $ \alpha  \beta - \delta \gamma = 1 $, and the matrix ${\bf L} $ was
defined earlier.
 Substituting the values
 
\beq\label{eq2.16}
\g  \quad = \quad \frac{1}{t} \, \pmatrix{2 & 1 \cr
  1 & 2 }
 \qquad \qquad
\b \quad = \quad \frac{1}{t} \, \pmatrix{0 & k'\cr -k'& 0  }  
\eeq
\vskip 0.6cm
\noindent
after some algebra one finds 
\beqa\label{eq2.17}
\g' \quad  & =  & \frac{1}{2}\quad   \pmatrix{2 g_{11}' & g_{11}'\cr
  g_{11}' & 2 g_{11}' } \cr
&&\cr
&&\cr
g_{11}' \quad & =  & \quad  \frac{4 \, t } { 2 \, \alpha^2 \, t^2 + 2 \,
  \gamma^2
 (k'^2 + 3) - 4\, \alpha\,\gamma\,k'\,t }
\eeqa
\noindent
which yields the most general $PSL(2,Z) $ transformed  filling fraction
 
\beq\label{eq2.18}
\nu' \quad = \quad  \frac{2}{ \alpha^2 \, t  - 2 \, \alpha  \, \gamma  \, k' + 
\frac{{\displaystyle \gamma}^2}{\displaystyle t}  ( k'^2 + 3 ) }
\eeq
\vskip 0.6cm
\noindent
What remains is how to interpret the (quasi)-particles in the edge theory
that  are responsible for these new fractions. The situation is
simplified by the fact that conditions $(i) $ and $(iii) $
in eq(\ref{eq1.21}) are satisfied  by  $(\b', \g' ) $  if  they are
satisfied by the initial couplet 
$(\b , \g ) $. (In fact the spin $N^t  \, M $ is  duality invariant ). This
means that the quasi-particles  of the transformed fractions will have Fermi
 statistics if  we take the  particular solution for $(\b , \g ) $ 
given in eq(\ref{eq2.9}). The issue concerning what charges they have is
more subtle as the condition $M_{L1}$ is not invariant but transforms
under $PSL(2,Z)$. In general we can expect that the charges of the
transformed states are fractions of  $-e$ . This fact has been highlighted
in [15], and a priori it may not be unacceptable that certain filling
fractions are described by such particles. Having said that,  we can try
and  see if  any of the fractions 
given in eq(\ref{eq2.18})  can correspond to electronic edge states. In
the toroidal theory, new fractions were obtained that did have this interpretation.
 The "trick" in showing this was to note that if one has  
as a result of a  particular duality transformation, $\b' \, = \, 0 $, then
 conditions $(ii) $ and $(iii) $ are equivelant,  and the corresponding
 couplet will be a solution. In the $T^2 $ case, a representitive  
$O(2,2;Z) $ transformation that achieves this is (in  our notation)

\beq\label{eq2.19} 
\b \quad \stackrel{{\displaystyle \omega}'}{\longrightarrow}\quad \b' \quad = \quad 0 ,
\qquad 
 \omega' \, = \, 
\pmatrix{{\bf I}_2&- {\bf L} \cr 0& {\bf I}_2 }
\eeq
\vskip 0.6cm
\noindent
with the initial couplet

\beq\label{eq2.19b} 
 \g \,\,  = \,  \,  \frac{1}{2mp} \, \pmatrix{ 2p & -1\cr  -1 & m }, \quad \quad 
\b \,\,  = \, \, \frac{1}{2 mp} \, \pmatrix{0&1\cr -1&0 } 
\eeq
\vskip 0.6cm
\noindent
Returning to our example, the most general transformation of the
combination
 $ \b {\g}^{-1} $ 
under $PSL(2,Z) $ is
 
\beq\label{eq2.20}
\b' \, \g'^{-1} \, = \,  \frac{t}{3} \, \pmatrix{   \beta \gamma 
  \frac{( b^2
  t^2 +3 )}{{\displaystyle t}^2} - \alpha \beta b
+\alpha \delta  -  \gamma \delta  \, b & -2\beta \gamma  \frac{( b^2  t^2
  +3 )}
{{\displaystyle t}^2} +2 \alpha \beta b
-2 \alpha \delta  + 2   \gamma \delta   b \cr&\cr
2\beta \gamma  \frac{( b^2  t^2 +3 )}{{\displaystyle t}^2} -2 \alpha \beta b
+2 \alpha \delta  - 2   \gamma \delta   b & -\beta \gamma  \frac{( b^2
  t^2 +3 )
}{{\displaystyle t}^2} + \alpha \beta b
-\alpha \delta  +  \gamma \delta   b}
 \eeq
where  we remind the reader that  $b = k'/t $.  We can determine
$\b' $ from eq(\ref{eq2.20})
given the form of $\g' $  in eq(\ref{eq2.17}). 

If we now consider the solutions to the  condition  $\b' = 0 $, then we
find two constraints
 
\beq\label{eq2.21}
 \qquad \qquad  \beta \, \gamma \, = \, 0  \qquad , \qquad  \alpha \, \delta -
 ( \beta \, \alpha + \gamma \, \delta ) \, b \, = \, 0
\eeq
for which there are two solutions
\beqa\label {eq2.22}
(i) \qquad \qquad   \gamma \,  & =  & \, 0\qquad ,  \qquad \alpha \, = \, \beta \, =
\pm 1 , \qquad  b \, =  \, \pm \delta \cr
&& \cr
(ii) \qquad \qquad  \beta \, &  =  &\, 0 \qquad , \qquad \gamma \, = \, - \delta \, =
\, \pm 1 , \qquad  b \, = \, \pm \alpha 
\eeqa
\vskip 0.6cm
\noindent
These solutions imply that in $(i) \quad k' \, = \, \pm \delta \, t $ and
$(ii) \quad k' \, = \, \pm\alpha \, t $
respectively, which means that $k' \, \in \, 2\, Z $. However
this contradicts the assumption that  $k' \, \in \, 2 \, Z + 1 $
 which was required to ensure that the edge state described
fermionic  particles (see below eq(\ref{eq2.11})). Thus  in order to
maintain charge $-e $ particles in the edge  theory  ( setting $\b' \, = \,
0 $ being one way of achieving this) we are forced to have bosonic 
edge states. Conversely if we demand that our original  couplet $(\b , \g )
$ describe charge $-e$ fermions, 
then we have to accept that in the transformed theory $\b' \neq 0 $, and
the fermionic states in this theory 
will have fractional charges in general. This situation is clearly very
dependent on the particular choice of 
orbifold model, and given the huge class of orbifolds that have appeared in
the string literature (see
[31] for a partial classification) it is very hard to imagine that models
cannot be found where we can 
have charge $-e $ fermionic edge states in both the original and
transformed theories, as in the toroidal case. Work in obtaining examples
of this kind is in progress [25]. 

So let us continue with keeping  our general transformed $( \b' , \g' ) $
and calculate the  general expression for the transformed electric charge
$q' $ of the fermionic  edge states that give the filling fractions
eq(\ref{eq2.18}) .  Using eqs(\ref{eq1.22}) and the transformation
properties of $M_{Li} $ 
the result is 
\beq\label{eq2.23}
q' \, = \,  e \, \frac{2 \, t \, (- \alpha \, t - \gamma \, ( 3 \, \rho - k'
  ) \, ) }{2 \, \alpha^2 \, t^2  - 4 \, \alpha
\, \gamma \, k' \, t  + 2\, \gamma^2 \, (k'^2 +3 ) } 
\eeq
\vskip 0.6cm
\noindent
There are some obvious  choice of elements $\omega \, \in  \, PSL(2,Z)
$ (other than the identity)
for which the transformed charge $q' $ is an integer multiple of $ e
$. Consider the ``axionic shift''
elements given by $\alpha \, = \, \beta \, = \, \pm 1 \,  , \, \gamma
\, = \, 0  $ and  $\delta \neq 0 $.
In this case $q' \, = \, q \, =  \,-e $ since these shifts do not affect
the charge, and furthermore they also 
leave the filling fraction invariant (its easy to see from
 eq(\ref{eq2.18}) that  $ \nu' \, = \, \nu $ for these shifts.)  It is not obvious whether
these are the only elements for which $q' $ is an integer multiple of $e $, 
As we anticipated, $q' $ given in eq(\ref{eq2.23}) is
generically  given by fractional multiples of the electronic charge.
Finally we  look at  some particular $PSL(2,Z) $  elements
and the filling fractions they give. 
Taking   for example, $\alpha \, = \, 0  \, , \quad \gamma \, = \, -\delta \, = \, \pm 1  $
gives the  filling fraction 
and charge
 
\beq\label{eq2.24}
\nu' \, = \, \frac{2\, t}{3 + x^2 } \qquad, \qquad 
\frac{q'}{e} \, = \, \frac{ \pm  t \, (3 \, \rho - x  ) }{x^2 +3 }
 \eeq
\vskip 0.6cm
\noindent
where in obtaining these equations we have set, without loss of generality,
$b \, =  \, k'/t \, =  \,  (\alpha \, t + x)/t \quad , \quad
x \, \in \, 2 \, Z + 1  $, the latter condition on $x$ ensuring
that we have  fermionic edge states.
Recall that $\rho \, \in \, 2\, Z $, and $m \, \in \, 2\,
Z +1 $. 
Now taking $x\, = \, 1 $  gives   odd integer values of $\nu' \, = \, m $ 
 corresponding to the  integral quantum hall effect, since  $t \, = 2\, m
 $. Moreover, since we saw earlier that $m $ is required to be an odd
 integer divisible by $3$, another choice  $ x\, = \,  3  $ gives arbitrary
 odd integer filling fractions 
$\nu' \, = \, m/3 $. In both these cases the corresponding  charges 
 are given by odd integer 
multiples of $-\frac{1}{2} \, e $. Unfortunately, quasi-particles carrying 
fractional charges with 
even denominators are apparently ruled out by physical data, so the $Z_3 $
orbifold toy-model under discussion in this section must remain just that.
In any case it would be somewhat remarkable if one of the simplest 2-dimensional orbifolds
should satisfy all the physical requirements. At least 
the model illustrates  the existence
of duality transformations that 
connect the original filling fraction $\nu \, = \,
  \frac{1}{\displaystyle m} $ with
states with $\nu' \, =\, m $ and $\frac{\displaystyle m}{\displaystyle 3} $,
 each case corresponding to
a different choice of $b $ field. It also provides positive  motivation
in searching for other orbifold models wich yield more realistic quasi-particle   
spectra.

Another  notable set of  filling fractions are those of the Jain hierarchy
[3],  which are given by the series
 \beq\label{eq2.25}
\nu \, = \, \frac{1}{2 \, p + \frac{ 1}{\displaystyle m}} \qquad  , \qquad  \nu \, = \,
\frac{1}{2 \, p - \frac{1}{\displaystyle m}}
\eeq
\vskip 0.6cm
\noindent
with $m \in 2 \, Z + 1  \, , \,   p \, \in \, Z $ . From the general form
of the transformed  fraction 
eq(\ref{eq2.18}), taking $k' \, = \, 1\, , \alpha \, = \, \beta \, = \,
 +1,   \,  \gamma \, = \, \pm 1 \,  ,
\, \delta \, = \, 0  \, $ gives  $ \nu' \, = \,   ( m \pm 1 + \frac{1}{m}
{)}^{-1} $
 which is a Jain fraction with  $p = \frac{ \displaystyle m \pm 1 }{\displaystyle 2} $.  Similarly
 another subset of Jain fractions can be obtained from the 
values  $k' \, = \, 3\, , \alpha \, = \, \beta \, = \,  +1,  \,
  \gamma \, = \, \pm 1 \,  ,
\, \delta \, = \, 0  \, $ gives  $ \nu' \, = \,   ( m \pm 3 +
 \frac{\displaystyle 3}{\displaystyle m}
{)}^{-1} $  which  corresponds to 
$p\, = \, \frac{\displaystyle m \pm 3}{\displaystyle 2} $. These are Jain fractions by virtue of the
fact that the odd integer $m$ is divisible by  $3$ as we saw earlier.

 One may ask what are the most general Jain fractions one can obtain this
way. Writing $k' \,  \, = \,  2 w + 1 \, , \,  w \, \in \, Z  $,   then we can write 
\beq\label{eq2.26}
\nu' \, = \,  \frac{1}{ 2 \, p + \frac{1}{{\displaystyle m}'} }
  \quad , \quad  p \, \in \, Z \, ,  \qquad 
m' \, \in \, 2 \, Z + 1 
\eeq
if 
\beq\label{eq2.27}
m \, = \, m' \, \gamma^2 \, (  w^2 + w + 1)  \quad , \quad  2\, p\, =\, 
\alpha\, (\alpha m' \gamma^2 \, (w^2 +w +1 ) - \gamma \, ( 2 w +1 ) )  
\eeq
Since $m $ and $m' $ are both odd integers, then we require 
 $\gamma \, \in \, 2 \, Z + 1 $. 
Then one can easily see that the right hand side of the second  equation in
 eq(\ref{eq2.27}) is always an even integer, for arbitrary integer values of 
$\alpha $ and $w$, and arbitrary odd integers $\gamma$ , subject of course
to the constraint $\alpha \, \beta - \gamma \, \delta \, = \, 1 $. This therefore
defines the integer $p$ of the Jain fraction. Whether eqs(\ref{eq2.26}) and
(\ref{eq2.27}) constitute a complete parameterization of Jain fractions of
the type given in the first part of eq(\ref{eq2.26}) is an interesting
question. Certainly the situation in the $Z_3 $ orbifold case is different
than in the $T^2 $. There by taking different
starting values for $\g $ and $\b $, it was shown that  the Jain
fractions eq(\ref{eq2.26}) can be obtained via certain $O(2,2;Z) $
transformations of the original fraction
 $\nu \, =\, \frac{1}{\displaystyle m} $. Since
such dualities are a symmetry of the total edge state spectrum, one of the
surprising conclusions in [20] is that the fraction
 $\nu \, = \, \frac{1}{\displaystyle m}
$ and those of the Jain hierarchy, share the same spectrum.

In our case the connections are more subtle and complex, in particular 
there will be different values of the odd integers $m$ and $m' $ 
 that enter the Haldane  and  Jain fractions, unlike the toroidal case
 where they are equal. Of course it has to be remembered that this
 is a toy  model, only illustrative of the possible applications of
 orbifold edge theories as remarked earlier.

\section {Conclusions}
In this paper we have presented a general formalism for investigating the
possible role that chiral orbifold CFT and their corresponding duality
symmetries, may play in understanding the interconnection between various
quantum Hall hierarchies. Considering orbifold edge theories in this way,
provides a natural generalization of the previous work in the literature,
which has made use of the well known toroidal duality group $O(n,n;Z ) $
present in strings compactified on $T^n $. Strings compactified on
orbifolds share many features with those compactified on $T^n $ as well as
 introducing new features. One of these features is to place further
 geometric conditions on the allowed values of the background fields $G_{ij}
 $ and $B_{ij} $ which affects directly the allowed values of Hall
 conductivity $\sigma_H $ and hence the filling fractions one can obtain by
 applying duality transformations to the couplet $ ( \g , \b ) $.
We have derived the general conditions, analogous to the  toroidal edge
theory, which imply that the states responsible for  these
 filling fractions are fermions carrying charge $-e $. 

The connection  between  chiral orbifold edge theories and  Luttinger liquids,
and their CFT representation in terms of  $n$-flavour massless Thirring models 
was also investigated.  The geometric constraints placed on the background
fields $\g  \, , \,  \b $ manifest themselves as  constraints on the  Thirring model
couplings  $h_{ij} $ . These in turn should  place constraints on the  various couplings 
of the $n$-flavour Luttinger model. It would be interesting to investigate the details of this,
as one would then learn how the geometric symmetries of orbifolds directly translate into
"selection rules"  for allowed interactions in the generalized Luttinger model.
At the same time, it would be worthwhile  investigating charge transport  between 
boundaries in this model, as this would provide an
alternative  derivation of  the kinds of filling fractions one can find in  orbifold edge theories.    

We  illustrated the general formalism presented in this paper, with a simple toy model based on 
the 2-dimensional $Z_3 $ orbifold, and have seen that the basic constraints
allow a description of Haldane fractions with
 $\nu \, =  \, \frac{1}{\displaystyle m}$,  $\quad m$
odd. The most general duality transformed filling fraction $\nu ' $ was
derived and although this toy model has the feature that the corresponding
transformed edge states carry fractional electronic charges in general, 
we have seen that integer filling fractions as well as certain subset of 
Jain's hierarchy can be obtained this way.

As we have mentioned earlier, the fact that these latter states have
particular fractional charges that are apparently ruled out by physical
data  is a consequence of the specific  $Z_3 $  orbifold in this
example, and is not a feature that is expected to hold for all
orbifolds. Indeed since the known class of orbifolds is vast, and their
properties quite diverse, we anticipate that
a systematic search should yield  models with more physical quasi-particle spectra. 

\vspace{6ex}
\begin{center}
{\bf\Large Acknowledgments}
\end{center}
\vskip 0.8cm
S. Skoulakis would like to thank the Onassis foundation and PPARC  for
funding, and S. Thomas the Royal Society for financial support.

\vspace{8ex}
\begin{center}
{\bf\Large References}
\end{center}
\vspace{2ex}
\begin{description}
\item{[1]} R.E. Prange and S.M. Girvin, eds., { \it  The Quantum Hall Effect}
                  Springer-Verlag, New York (1990)(second ed.)  and
                  references therein; 
           DasSarma Sanker ``Perspectives in Quantum Hall Effects:
           Novel Quantum Liquids in Low-Dimensional Semiconductor
           Structures'', Wiley, New York, (1996) (and references therein);
           M. Jonssen, O. Viehweg, U. Fastenrath and J. Hajdu, 
           `` Introduction to the Theory of the Integer Quantum Hall
           Effect'' New York : VCH (1994) (and references therein).
\item{[2]} X.G. Wen, {\it Topological orders and edge excitations in FQH
            states}, cond-mat/9506066
\item{[3]} J.K. Jain, {\it Adv. phys.}  { \bf  41}  (1992) 105
\item{[4]} J. Frohlich and T. Kerler {\it Nucl. Phys.} {\bf B354} 
                   (1991); J. Frohlich and U.M. Studer, Rev. Mod. Phys. {\bf 65}, 733
                    (1993); J. Frohlich and E. Thiran (Zurich THH 93-22) 
                    ; J. Frohlich, U.M. Studer, E. Thiran in {\it Les Houches
                    Summer Scholl Session 62 : Fluctuating  Geometries in
                    Statistical Mechanics and Field Theory } Les Houches
                    France 2 Aug. -9 Sept  1994, cond-mat/9508062
\item{[5]} C.A. Lutken and G.G. Ross, {\it Phys. Rev.}  { \bf B45} (1992) 11837 ; 
           {\it Phys. Rev.}  { \bf B48} (1993) 2500
\item{[6]} B. P. Dolan, {\it Duality and the Modular group in Quantum Hall
              Effect } cond-mat/9805171
\item{[7]} E. Witten, {\it Comm. Math. Phys.} {\bf 121} (1989) 351; S. Elitzur, G, Moore, 
           A.  Schwimmer and  N. Seiberg, {\it Nucl. Phys.}  {\bf  B326} (1989) 108 
           and references therein.
\item{[8]} G. Moore and N. Seiberg {\it Phys. Lett } { \bf B220}  (1990) 422
\item{[9]} A.P. Balachandran, G. Bimonte, K.S. Gupta and A. Stern,  {\it Int. J. Mod. Phys}
           {\bf A7} 4655  and 5855 (1992)
\item{[10]} F.D.M. Haldane, {\it J.Phys.} {\bf C}:Solid State Phys. { \bf 14} (1981)
            2585 
\item{[11]} C.G. Callan, Jr. And J.A. Harvey, {\it Nucl. Phys.} {\bf  B250} (1985) 427.
\item{[12]} S.G. Naculich, {\it Nucl. Phys} {\bf  B296} (1988) 837.
\item{[13]} F. Wilczek,  "Fractional Statistics and Anyon Superconductivity" 
                   (World Scientific press 1990) ; M. Stone, {\it
                     Ann. Phys.}  { \bf 207} (1991) 38
\item{[14]} A. Shapere and F.Wilczek, {\it Nucl. Phys.} {\bf B320} (1989) 669; 
            A. Giveon, E. Rabinovici and G. Veneziano,  {\bf  Nucl. Phys.} {\bf B322} 
           (1989)  167; A. Giveon and M. Rocek, {\it Nucl. Phys.} {\bf B380} (1992)
          128; J. Maharana and J.H. Schwarz, {\it Nucl. Phys.} {\bf B390} (1993) 3
          A.Giveon, M. Porrati and E. Rabinovici, {\it Phys. Rep. } {\bf
          244 } (1994) 77
\item{[15]} A.P. Balachandran, L. Chandar, B. Sathiapalan, {\it Nucl. Phys.} {\bf B443}  (1995)  465 
\item{[16]} L. Dixon, J.A. Harvey, C. Vafa and E. Witten, {\it Nucl. Phys.}
           {\bf B261 } (1985) 678 ; {\it Nucl. Phys.} {\bf B274 } (1986) 285
\item{[17]} P. Degiovanni, C. Chaubet, R. Melin, { \it  Conformal field theory 
           approach to gapless 1D fermion systems and application to the edge
           excitations of}  $ \nu  = \frac{1}{\displaystyle  2p+1 } $ {\it quantum Hall sequences.}
           cond-mat/9711173  
\item{[18}]  M. Spalinski {\it Phys. Lett }{\bf B 275 } (1992) 47 ; J.
            Erler, D. Jungnicker and H.P. Nilles, {\it  Phys. Lett }{\bf B 276 }
            (1992) 303
\item{[19]} A. Love, W.A.  Sabra and  S. Thomas, {\it Nucl. Phys} { \bf  B427} (1994) 181
\item{[20]} K.S. Narain {\it Phys. Lett.}  {\bf B 169} (1987)  41;
            K.S. Narain, M.H. Sarmadi and E. Witten, {\it  Nucl. Phys.} { \bf  B279} (1987) 369
\item{[21]} X.G. Wen {\it Phys. Rev. } {\bf B411} (1990) 12838.
\item{[22]} S.Thomas, {\it Int. J. Mod. Phys. } {\bf A4} (1989) 2561 
\item{[23]} P. Mayr and  S. Steiberger,  {\it Nucl. Phys.} { \bf B407}
            (1993) 725
\item{[24]} D. Bailin, A. Love, W. Sabra and S. Thomas,  {\it Mod. Phys.
             Lett. } {\bf A9 } (1994) 67 ; {\it Phys. Lett. } {\bf B320 } (1994) 21
\item{[25]} S. Skoulakis and S. Thomas, work in progress.
\item{[26]} I.I. Kogan, {\it Mod. Phys. Lett. } {\bf A6} (1991) 501;
\item{[27]} L. Copper, I.I. Kogan and K.M. Lee {\it Phys. Lett. } {\bf
            B 394} (1997) 67 
\item{[28]} L. Copper, I.I. Kogan and R.J. Szabo, {\it Mirror Maps in
            Chern-Simons gauge theory } OUTP-97-57P, hep-th/9710179
\item{[29]} J. Bagger, N. Seiberg, S. Yankielowicz, {\it  Nucl. Phys.}  {
             \bf B289} (1987) 53  
\item{[30]}  J. Cardy, {\it Nucl. Phys.} {\bf B270} (1986), 186.
\item{[31]} Y. Katsuki, Y. Kawamura, T. Kobayashi, N. Ohtsubo, K. Tanioka, 
           {\it Nucl. Phys.} { \bf B341} (1990) 611
\item{[32]} A. Cappelli and  G.R. Zemba,  {\it Nucl. Phys. }  { \bf  B490}
            (1997)  595.
\item{[33]} I.E. Dzyaloshinskii and A.I. Larkin {\it Sov. Phys. JETP }
             {\bf 38} (1974) 202
\item{[34]} J.M. Luttinger, {\it J. Math. Phys.} {\bf 4} (1963) 1154.
\item{[35]} A.H. MacDonald in {\it Mesoscopic Quantum Physics, Les houches}1994 session LXI.
\item{[36]} H.J. Schulz in {\it Mesoscopic Quantum Physics, Les houches}1994 session LXI.

\end{description}

\end{document}